# On the Schrödinger Functional in QCD

Stefan Sint

Deutsches Elektronen-Synchrotron DESY
Notkestrasse 85, D-22603 Hamburg, Germany

**Abstract**

In a series of publications [1,2], Lüscher et al. have demonstrated the usefulness of the Schrödinger functional in pure SU(2) and SU(3) gauge theory. In this paper, it is shown how their formalism can be extended to include fermions. In the framework of Wilson's lattice QCD, we define the Schrödinger functional by making use of the transfer matrix formalism. Boundary conditions for the fermions arise naturally. We then take the naive continuum limit of the action and show that no lattice peculiarities are left over. The corresponding free Dirac operator has a unique self-adjoint extension with purely discrete spectrum and no zero modes.

# 1. Introduction

Recently, it has been shown how the Schrödinger functional can be used for a theoretical determination of the running coupling in asymptotically free gauge theories [1]. The subsequent applications to the SU(2) and SU(3) Yang-Mills theories have been very successful [2], so that one may feel encouraged to extend this framework to QCD.

The Schrödinger functional as defined in reference [1] is the euclidean propagation kernel for going from an initial field configuration at time $x_0 = 0$ to a final configuration at euclidean time $x_0 = T$. Following Feynman, such a transition amplitude can be written as a path integral with the usual (euclidean) action, where the integral runs over all interpolating field configurations. The quantum fields then fluctuate around the "classical path", i.e. a classical background field, which may be modified by tuning the initial and final field configurations.

The Schrödinger functional is important for the application mentioned above, since it allows to define a running coupling as the system's response to the induced background field. This definition is not inherently based on the perturbation expansion and is therefore also well-suited for computations in the non-perturbative low-energy domain.

It is the purpose of the present paper to show how the Schrödinger functional can be defined in QCD. In a theory with fermions, it is not clear what one should take as initial and final field configurations. In the euclidean framework this amounts to the problem of finding adequate boundary conditions for the quark fields. The difficulty arises because the Dirac equation is only of first order, and therefore, one can only prescribe the values of half of the spinor components at the boundary.

To clarify the situation, it is most convenient to define the Schrödinger functional on the lattice, using the transfer matrix formalism. The lattice action and the boundary conditions then emerge naturally. As a first check, we then take the continuum limit of the lattice action and verify whether this leads to an acceptable classical continuum theory.

The paper is organized as follows. Section 2 introduces some of our conventions. The case of pure gauge theory is reviewed in section 3, stressing the rôle of the transfer matrix. We then introduce the transfer matrix for Wilson fermions and discuss a convenient technique for obtaining its Grassmann representation (sect.4). The Schrödinger functional is defined and the lattice action and the boundary conditions are read off from its path integral representation. Sects. 5 and 6 are concerned with the classical theory, and we end with a few remarks concerning questions of immediate interest.

## 2. Basic conventions

In order to introduce some conventions and notations, we consider QCD on a finite hyper cubic euclidean lattice

$$\Gamma_{\mathrm{E}} = \left\{ x \,\big|\, x/a \in \mathbb{Z}^4,\ 0 \leq x^0 < T,\ 0 \leq x^k < L,\ k = 1, 2, 3 \right\}. \tag{2.1}$$

A classical quark field is an assignment of a Dirac spinor $\psi(x)$ with components $\psi(x)_{sa\alpha}$ to each lattice site. The indices $s$ and $a$ denote the flavour and colour degrees of freedom and $\alpha$ is the Dirac index. A gauge field is an ensemble of matrices $U(x,\mu) \in \mathrm{SU}(N)$, residing on the bonds $(x, x + a\hat{\mu})$ of the lattice ($\hat{\mu}$ denotes a unit vector in $\mu$-direction). Throughout the paper, periodic boundary conditions in the spatial directions are assumed for all fields. Assuming periodicity in the time direction too, one may write down the action, $S = S_g + S_f$, according to Wilson [3] as follows

$$S_g[U] = \frac{1}{g_0^2} \sum_{x \in \Gamma_{\mathrm{E}}} \sum_{\mu,\nu=0}^{3} \mathrm{tr}\{1 - P_{\mu\nu}(x)\}, \tag{2.2}$$

$$S_f[\psi, \bar{\psi}, U] = a^3 \sum_{x \in \Gamma_{\mathrm{E}}} \left\{ \bar{\psi}(x)\psi(x) - \sum_{\mu=0}^{3} [\bar{\psi}(x)U(x,\mu)\mathcal{K}(1 - \gamma_\mu)\psi(x + a\hat{\mu}) \right.$$
$$\left. + \bar{\psi}(x + a\hat{\mu})U(x,\mu)^{-1}\mathcal{K}(1 + \gamma_\mu)\psi(x)] \right\}. \tag{2.3}$$

Here, $P_{\mu\nu}(x) = U(x,\mu)U(x + a\hat{\mu}, \nu)U(x + a\hat{\nu}, \mu)^{-1}U(x,\nu)^{-1}$ is the usual plaquette field and $\mathcal{K}$ is the hopping parameter matrix, assumed to be diagonal and non-negative, its relation to the bare quark mass matrix $m$ being $\mathcal{K}^{-1} = 2ma + 8$. Our $\gamma$-matrices are hermitian and satisfy

$$\{\gamma_\mu, \gamma_\nu\} = 2\delta_{\mu\nu}. \tag{2.4}$$

In the continuum one is used to a different normalization of the spinor fields, namely $\psi_c = (2\mathcal{K})^{1/2}\psi$ ($c$ stands for continuum). Later on, we will consider the continuum theory and then use this normalization. If one introduces the covariant lattice derivatives

$$\begin{aligned} \nabla_\mu \psi(x) &= \frac{1}{a}\left[ U(x,\mu)\psi(x + a\hat{\mu}) - \psi(x) \right], \\ \nabla_\mu^* \psi(x) &= \frac{1}{a}\left[ \psi(x) - U(x - a\hat{\mu}, \mu)^{-1}\psi(x - a\hat{\mu}) \right], \end{aligned} \tag{2.5}$$

and the difference operators $\tilde{\nabla}_\mu = (\nabla_\mu + \nabla_\mu^*)/2$ and $\Delta = -\sum_{\mu=0}^{3} \nabla_\mu^* \nabla_\mu$, one can rewrite the fermionic action $S_f$ in the form

$$S_f = a^4 \sum_{x \in \Gamma_{\mathrm{E}}} \bar{\psi}_c(x) \left\{ \sum_\mu \tilde{\nabla}_\mu \gamma_\mu + m + \frac{1}{2}a\Delta \right\} \psi_c(x), \tag{2.6}$$

which is reminiscent of the euclidean action in the continuum.

# 3. Pure SU(N) gauge theory reviewed

The Schrödinger functional in pure SU($N$) gauge theory has been discussed in reference [1], where also a formal approach in the continuum can be found. We find it useful to go through the lattice definition of this functional in some detail. First of all, it is evident that the pure gauge theory functional will constitute an integral part of the full QCD functional. Furthermore, from the discussion of the pure gauge theory, it will become clear how to proceed for lattice QCD.

The starting point is the quantum mechanics of the pure gauge theory. One considers $L$-periodic gauge fields on a spatial lattice $\Gamma$, which can be obtained as a sublattice of the euclidean lattice $\Gamma_E$ (2.1) for a fixed value of $x_0$ ($x_0 = 0$ for instance). Wave functions are defined as the complex-valued square integrable functions of the gauge fields. They form a Hilbert space with the scalar product given by

$$\langle \varphi_1 | \varphi_2 \rangle = \int \mathcal{D}[U] \varphi_1^*(U) \varphi_2(U), \qquad \mathcal{D}[U] = \prod_{\mathbf{x} \in \Gamma} \prod_{k=1}^{3} dU(\mathbf{x}, k). \tag{3.1}$$

The wave function $\varphi$ is the Schrödinger representation of a state vector $|\varphi\rangle$. It can be obtained from the latter by introducing a complete set of states $|U\rangle$, labelled by classical gauge fields $U$ such that

$$\langle U | \varphi \rangle = \varphi(U). \tag{3.2}$$

Since the field operators $\hat{U}(\mathbf{x}, k)$ are diagonal in the Schrödinger representation, the states $|U\rangle$ are required to be eigenstates of $\hat{U}(\mathbf{x}, k)$ with the corresponding classical field as eigenvalue. Completeness implies the decomposition of the identity operator

$$\mathbb{1} = \int \mathcal{D}[U] |U\rangle\langle U|, \tag{3.3}$$

which allows to retrieve a state vector from its Schrödinger representation.

Only gauge invariant wave functions, which satisfy $\varphi(U) = \varphi(U^V)$ are physical. Here, $U^V$ denotes the gauge transformed field $U^V(\mathbf{x}, k) = V(\mathbf{x}) U(\mathbf{x}, k) V(\mathbf{x} + a\hat{k})^{-1}$, with $V(\mathbf{x})$ being an $L$-periodic SU($N$)-matrix field on $\Gamma$. Any wave function can be projected on the physical subspace with the help of the projector $\mathbb{P}$,

$$\mathbb{P}\varphi(U) = \int \prod_{\mathbf{x} \in \Gamma} dV(\mathbf{x}) \, \varphi(U^V). \tag{3.4}$$

The (euclidean) time evolution of a wave function is now provided by the transfer matrix $\mathbb{T} = \mathbb{T}_0 \mathbb{P}$, which can be thought of as the euclidean step evolution operator $e^{-\mathbb{H}a}$ (with Hamiltonian $\mathbb{H}$). The operator $\mathbb{T}_0$ acts on wave functions as an integral operator

$$\mathbb{T}_0 \varphi(U) \equiv \langle U | \mathbb{T}_0 | \varphi \rangle = \int \mathcal{D}[U'] \langle U | \mathbb{T}_0 | U' \rangle \varphi(U'), \tag{3.5}$$

with kernel $K_0[U, U'] \equiv \langle U|\mathbb{T}_0|U'\rangle = \exp -\Delta S[U, 1, U']$. When making contact with the path integral formulation, one usually imagines a double layer of two space lattices $\Gamma$, separated by a euclidean time $a$. Apart from the factors $1/2$, the function $\Delta S[U, V, U'] = S_p[U]/2 + S_k[U, V, U'] + S_p[U']/2$ then equals the euclidean action of this double layer, provided $V$ is interpreted as the temporal component of the euclidean gauge field. For the Wilson plaquette action (cf. sect.1), one defines

$$S_p[U] = \frac{1}{g_0^2} \sum_{\mathbf{x}\in\Gamma} \sum_{k,l=1}^{3} \text{tr}\{1 - P_{kl}(\mathbf{x})\},$$
$$S_k[U, V, U'] = \frac{1}{g_0^2} \sum_{\mathbf{x}\in\Gamma} \sum_{k=1}^{3} \text{tr}\{2 - P_{k0}(\mathbf{x}) - P_{0k}(\mathbf{x})^{-1}\}. \tag{3.6}$$

For obvious reasons, the operator $\mathbb{T}_0$ is referred to as transfer operator in the temporal gauge. Using the gauge invariance of the action,

$$\Delta S[U, V, U'] = \Delta S[U, 1, U'^{V^{-1}}] = \Delta S[U^V, 1, U'], \tag{3.7}$$

and the invariance of the measure, one easily sees that $\mathbb{T}_0$ commutes with $\mathbb{P}$ and the integral kernel of the transfer matrix $\mathbb{T} = \mathbb{T}_0\mathbb{P}$ is given by

$$\langle U|\mathbb{T}|U'\rangle = \int \prod_{\mathbf{x}\in\Gamma} dV(\mathbf{x})\, e^{-\Delta S[U, V, U']}. \tag{3.8}$$

The rôle of the projector $\mathbb{P}$ therefore is to supply the temporal components of the euclidean gauge field.

Euclidean time transport over a time $T = Na$ is now obtained by iterating the action of the transfer operator $N$ times, viz.

$$(\mathbb{T}^N \varphi)(U_N) = \int \mathcal{D}[U_0]\, \langle U_N|\mathbb{T}^N|U_0\rangle \varphi(U_0). \tag{3.9}$$

The kernel can be evaluated by inserting the identity (3.3) $N - 1$ times,

$$\langle U_N|\mathbb{T}^N|U_0\rangle = \int \prod_{l=1}^{N-1} \mathcal{D}[U_l]\, \langle U_N|\mathbb{T}^N|U_{N-1}\rangle \cdots \langle U_2|\mathbb{T}|U_1\rangle\langle U_1|\mathbb{T}|U_0\rangle. \tag{3.10}$$

If one inserts the expression (3.8) and identifies the index $l$ with the euclidean time $x_0/a$ one obtains the path integral representation of the integral kernel

$$\langle W'|\mathbb{T}^N|W\rangle = \int D[U]\, e^{-S[U]}, \tag{3.11}$$

where the integration is over the euclidean gauge fields $U(x, \mu)$ with $0 \leq x_0 < T$ for $\mu = 0$ and $0 < x_0 < T$ for $\mu = 1, 2, 3$. The action $S$, given by

$$S[U] = \frac{1}{g_0^2} \sum_{\mathbf{x} \in \Gamma} \sum_{x^0=0}^{T} \sum_{\mu,\nu} w(P) \mathrm{tr}\{1 - P_{\mu\nu}(x)\}, \qquad (3.12)$$

is the usual Wilson plaquette action if it were not for the weight factor $w(P)$. This factor is set to 1 for all plaquettes $P$ except the space-like ones at the boundary ($x_0 = 0, T$) where it is set to $1/2$. Furthermore, the euclidean gauge field is subject to the boundary conditions

$$U(x, k)|_{x^0=0} = W(\mathbf{x}, k), \quad \text{and} \quad U(x, k)|_{x^0=T} = W'(\mathbf{x}, k). \qquad (3.13)$$

It is the integral kernel $\langle W'|\mathbb{T}^{T/a}|W\rangle$ which is referred to as Schrödinger functional. However, being interested in the continuum limit, we will again follow reference [1] and consider the Schrödinger functional as functional of the continuum gauge potentials $C_k(\mathbf{x})$ and $C'_k(\mathbf{x})$. The relation with the lattice boundary fields $W$ and $W'$ is then established through parallel transport along the lattice bonds, i.e.

$$W(\mathbf{x}, k) = \mathcal{P} \exp\left\{ a \int_0^1 \mathrm{d}t\, C_k(\mathbf{x} + a\hat{k} - ta\hat{k}) \right\}, \qquad (3.14)$$

and analogously for the primed fields (the symbol $\mathcal{P}$ means a path ordered exponential such that the fields at larger values of $t$ come first). The Schrödinger functional is then written as follows

$$\mathcal{Z}[C', C] \stackrel{\text{def}}{=} \langle W'|\mathbb{T}^{T/a}|W\rangle. \qquad (3.15)$$

## 4. Extension to QCD

In order to extend the definition (3.15) to QCD, one may use the same strategy as in the pure gauge theory case. This means, one starts with the transfer matrix, raises it to the $N$-th power and represents it as an integral operator, acting on suitable wave functions. The integral kernel of this operator will then be the Schrödinger functional one is looking for. The transfer matrix for lattice QCD with Wilson fermions has been constructed by Lüscher [4] some time ago, when he proved positivity of Wilson's lattice QCD. Except for some details and different notations, we will follow this approach.

*4.1 Transfer matrix*

The starting point is again the time zero Hilbert space which is now the product space of the pure gauge theory Hilbert space and a fermionic Fock space, generated by the action of a set of canonical operators $\hat{\chi}(\mathbf{x})_{sa\alpha}$, $\hat{\chi}^\dagger(\mathbf{x})_{sa\alpha}$ on the vacuum state $|0\rangle$. These operators reside on the sites of a spatial lattice $\Gamma$ with periodic boundary conditions. The flavour, colour and Dirac indices are the same as for the classical fields (cf. sect.2). By definition, one has

$$\{\hat{\chi}(\mathbf{x})_{sa\alpha}, \hat{\chi}^\dagger(\mathbf{y})_{tb\beta}\} = \delta_{st}\delta_{ab}\delta_{\alpha\beta}\delta(\mathbf{x}-\mathbf{y}). \tag{4.1}$$

In fact, it is an important observation that one should not impose canonical anti-commutation relations on the operators $\hat{\psi}$ and $\hat{\psi}^\dagger$ which correspond to the fields appearing in the classical action. Instead, these are related to the canonical operators through

$$\begin{aligned}
\hat{\psi}(\mathbf{x})_{sa\alpha} &= \sum_{\mathbf{y},t,b,\beta} B^{-1/2}(\mathbf{x},\mathbf{y})_{sa\alpha;tb\beta}\,\hat{\chi}(\mathbf{y})_{tb\beta}, \\
\hat{\psi}^\dagger(\mathbf{x})_{sa\alpha} &= \sum_{\mathbf{y},t,b,\beta} \hat{\chi}^\dagger(\mathbf{y})_{tb\beta}\,B^{-1/2}(\mathbf{y},\mathbf{x})_{tb\beta;sa\alpha},
\end{aligned} \tag{4.2}$$

and thus fulfill the anti-commutation relation

$$\{\hat{\psi}(\mathbf{x})_{sa\alpha}, \hat{\psi}^\dagger(\mathbf{y})_{tb\beta}\} = a^{-3}B^{-1}(\mathbf{x},\mathbf{y})_{sa\alpha;tb\beta}. \tag{4.3}$$

The matrix $B$ is at this stage only required to be hermitian and positive. When the connection with the euclidean formulation is established, one finds †

$$B(\mathbf{x},\mathbf{y})_{sa\alpha;tb\beta} = \delta_{\alpha\beta}\Big\{\delta_{st}\delta_{ab}\delta(\mathbf{x},\mathbf{y}) - \mathcal{K}_{st}\sum_{k=1}^{3}\big[U(\mathbf{x},k)_{ab}\delta(\mathbf{x}+a\hat{k},\mathbf{y}) \\ + U(\mathbf{y},k)_{ab}^{-1}\delta(\mathbf{y}+a\hat{k},\mathbf{x})\big]\Big\}. \tag{4.4}$$

The matrix $B$ can be obtained as the kernel of the difference operator $\hat{B} = 1 - 6\mathcal{K} - a^2\mathcal{K}\sum_{k=1}^{3}\nabla_k^*\nabla_k$, taking into account the periodic boundary conditions on the spatial lattice. It is a hermitian and positive matrix if all hopping parameters are smaller than $1/6$, which will be assumed in the following.

A general state $|X(U)\rangle$ of the Hilbert space evolves in euclidean time through the action of the transfer operator

$$\mathbb{T}_0|X(U)\rangle = \int \mathcal{D}[U']\mathbb{T}_0[U,U']|X(U')\rangle. \tag{4.5}$$

---

† Concerning the spatial $\delta$-functions, we distinguish between $\delta(\mathbf{x},\mathbf{y})$ with $\delta(\mathbf{x},\mathbf{x}) = 1$, and the 3-dimensional lattice $\delta$-function $\delta(\mathbf{x}-\mathbf{y}) = a^{-3}\delta(\mathbf{x},\mathbf{y})$.

With respect to the gauge field $\mathbb{T}_0$ is an integral operator whose kernel has the structure

$$\mathbb{T}_0[U, U'] = \hat{T}_F^\dagger(U) K_0[U, U'] \hat{T}_F(U') . \qquad (4.6)$$

The kernel $K_0[U, U']$ is the same as for the pure gauge theory [cf. eq.(3.5)], while the operator $\hat{T}_F(U)$ is given by

$$\hat{T}_F(U) = \det(2\mathcal{K}B)^{1/4} \exp(-\hat{\bar{\chi}}\tfrac{1}{2}(1 - \gamma_0)C\hat{\chi}) \exp(-\hat{\bar{\chi}}M\hat{\chi}) . \qquad (4.7)$$

The adjoint spinor is defined as usual by $\hat{\bar{\chi}} = \hat{\chi}^\dagger \gamma_0$ and the matrix $M$ is related to $B$ through $M = 1/2 \ln[B(2\mathcal{K})^{-1}]$. The anti-hermitian matrix $C$ is given by

$$C(\mathbf{x}, \mathbf{y})_{sa\alpha;tb\beta} = \delta_{st} \sum_{k=1}^{3} (\gamma_k)_{\alpha\beta} \frac{1}{2} [U(\mathbf{x}, k)_{ab} \delta(\mathbf{x} + a\hat{k}, \mathbf{y}) \\ - U(\mathbf{y}, k)_{ab}^{-1} \delta(\mathbf{y} + a\hat{k}, \mathbf{x})] , \qquad (4.8)$$

and is obtained as matrix kernel of $\hat{C} = a \sum_{k=1}^{3} \gamma_k \widetilde{\nabla}_k$.

In order to define the Schrödinger functional, one needs a representation of the transfer matrix as an integral operator acting on wave functions. It is well-known that any operator in Fock space may be represented as an integral operator in an associated Grassmann algebra. There is a nice technique to represent this isomorphism, based on so-called "Grassmann coherent states" (see e.g. [5]). To apply this technique, it is convenient to pass to a two-spinor representation and to choose a representation for the $\gamma$- matrices. This is only an intermediate step and at the end one may re-express everything in a way that does not depend on the particular representation, chosen as follows ($\tau_k$, $k = 1, 2, 3$ denote the Pauli matrices)

$$\gamma_0 = \begin{pmatrix} 1 & 0 \\ 0 & -1 \end{pmatrix} , \quad \gamma_k = i \begin{pmatrix} 0 & -\tau_k \\ \tau_k & 0 \end{pmatrix} . \qquad (4.9)$$

We will frequently use the projectors $P_\pm = \frac{1}{2}(1 \pm \gamma_0)$ which are diagonal in this representation and used to project on the two-spinor operators, defined by

$$\hat{\chi} = \begin{pmatrix} \hat{\eta} \\ \hat{\zeta}^\dagger \end{pmatrix} , \quad \hat{\chi}^\dagger = (\hat{\eta}^\dagger, \hat{\zeta}) . \qquad (4.10)$$

The matrices $B$ and $C$ have projections $\bar{B}$ and $\bar{C}$

$$B = \begin{pmatrix} \bar{B} & 0 \\ 0 & \bar{B} \end{pmatrix} , \quad C = \begin{pmatrix} 0 & -\bar{C} \\ \bar{C} & 0 \end{pmatrix} . \qquad (4.11)$$

In the two-spinor space we introduce barred greek indices $\bar{\alpha}, \bar{\beta}..$, which take the values 1 or 2. Classical spinor fields on the spatial lattice may be treated in parallel to the operator fields, i.e. the classical fields $\chi(\mathbf{x})$ and $\bar{\chi}(\mathbf{x})$ with

$$\chi(\mathbf{x}) = (B^{1/2}\psi)(\mathbf{x}), \qquad \bar{\chi}(\mathbf{x}) = (\bar{\psi}B^{1/2})(\mathbf{x}), \tag{4.12}$$

are decomposed as follows

$$\chi = \begin{pmatrix} \eta \\ \zeta^+ \end{pmatrix}, \quad \bar{\chi} = (\eta^+, -\zeta). \tag{4.13}$$

Whenever the four-component spinors $\chi$ and $\bar{\chi}$ appear together with the two-component spinors $\eta, \eta^+$ and $\zeta, \zeta^+$ within the same formula, this decomposition is tacitly assumed.

*4.2 Grassmann coherent states*

We proceed by introducing the states Grassmann coherent states $\langle \eta^+ \zeta^+ |$ and $| \eta \zeta \rangle$. These are Fock-states with Grassmann coefficients, such that the following holds

$$\begin{aligned}\hat{\eta}(\mathbf{x})_{sa\bar{\alpha}} | \eta \zeta \rangle &= | \eta \zeta \rangle \, \eta(\mathbf{x})_{sa\bar{\alpha}}, \\ \langle \eta^+ \zeta^+ | \, \hat{\eta}^\dagger(\mathbf{x})_{sa\bar{\alpha}} &= \eta^+(\mathbf{x})_{sa\bar{\alpha}} \langle \eta^+ \zeta^+ |, \end{aligned} \tag{4.14}$$

for all possible values of the indices (analogous relations hold for the operators $\hat{\zeta}$ and $\hat{\zeta}^\dagger$). It is always assumed that canonical operators and Grassmann variables commute with each other. The Grassmann coherent states are eigenstates of the canonical operators, with the corresponding classical fields as eigenvalues. It is not difficult to see that they are uniquely determined if the normalization conditions $\langle \eta^+ \zeta^+ | 0 \rangle = 1 = \langle 0 | \eta \zeta \rangle$ are imposed (assuming $\langle 0 | 0 \rangle = 1$). One then finds

$$| \eta \zeta \rangle = \sum_{J=0}^{\infty} \frac{(-1)^{J(J-1)/2}}{J!} \left[ a^3 \sum_{\mathbf{x},s,a,\bar{\alpha}} \left\{ \hat{\eta}^\dagger(\mathbf{x})_{sa\bar{\alpha}} \, \eta(\mathbf{x})_{sa\bar{\alpha}} + \hat{\zeta}^\dagger(\mathbf{x})_{sa\bar{\alpha}} \, \zeta(\mathbf{x})_{sa\bar{\alpha}} \right\} \right]^J |0\rangle, \tag{4.15}$$

and, similarly,

$$\langle \eta^+ \zeta^+ | = \langle 0 | \sum_{J=0}^{\infty} \frac{(-1)^{J(J-1)/2}}{J!} \left[ a^3 \sum_{\mathbf{x},s,a,\bar{\alpha}} \left\{ \hat{\eta}(\mathbf{x})_{sa\bar{\alpha}} \, \eta^+(\mathbf{x})_{sa\bar{\alpha}} + \hat{\zeta}(\mathbf{x})_{sa\bar{\alpha}} \, \zeta^+(\mathbf{x})_{sa\bar{\alpha}} \right\} \right]^J. \tag{4.16}$$

The Grassmann fields will become integration variables. When several integrations have to be carried out, we will need different sets of Grassmann variables which are associated to the same canonical operators. In this case we will use a prime or an index $l$ with the understanding that e.g. $\eta_l(\mathbf{x})_{sa\alpha}$ will be the eigenvalue of $\hat{\eta}(\mathbf{x})_{sa\alpha}$ when acting on the eigenstate $| \eta_l \zeta_l \rangle$. In the euclidean framework this label will then become the euclidean time index.

The isomorphism alluded to above takes a general Fock state $|X\rangle$ to its Grassmann equivalent, given by $\langle \eta^+\zeta^+|X\rangle$. This is the wave function we were looking for. It only depends on the crossed Grassmann variables and therefore, such wave functions are sometimes called "holomorphic states" in the literature [6].

Using the rules of Grassmann integration [7], one obtains the following decomposition of the identity operation in the Fock space

$$\mathbb{1}_{\text{Fock}} = \int \prod_{\mathbf{x}\in\Gamma} \prod_{s,a,\bar{\alpha}} \mathrm{d}\eta^+(\mathbf{x})_{sa\bar{\alpha}} \mathrm{d}\eta(\mathbf{x})_{sa\bar{\alpha}} \mathrm{d}\zeta^+(\mathbf{x})_{sa\bar{\alpha}} \mathrm{d}\zeta(\mathbf{x})_{sa\bar{\alpha}}$$
$$\times \exp\left[-a^3 \sum_{\mathbf{y}} \sum_{t,b,\bar{\beta}} \left\{ \eta^+(\mathbf{y})_{tb\bar{\beta}}\, \eta(\mathbf{y})_{tb\bar{\beta}} + \zeta^+(\mathbf{y})_{tb\bar{\beta}}\, \zeta(\mathbf{y})_{tb\bar{\beta}} \right\}\right] |\eta\,\zeta\rangle\langle\eta^+\zeta^+|\,. \tag{4.17}$$

It is this "completeness relation" which allows to recover a Fock state $|X\rangle$ from its wave function $\langle\eta^+\zeta^+|X\rangle$, thus proving that one really has to do with an isomorphism. As can be seen from eqs.(4.1) and (4.14), both, fermionic operators and the Grassmann fields have canonical dimension 3/2 and the normalization of the Grassmann integrals is chosen as follows

$$\int \prod_{\mathbf{x}\in\Gamma} \prod_{s,a,\bar{\alpha}} \mathrm{d}\eta^+(\mathbf{x})_{sa\bar{\alpha}} \mathrm{d}\eta(\mathbf{x})_{sa\bar{\alpha}} \mathrm{d}\zeta^+(\mathbf{x})_{sa\bar{\alpha}} \mathrm{d}\zeta(\mathbf{x})_{sa\bar{\alpha}}$$
$$\times \exp\left[-a^3 \sum_{\mathbf{y}} \sum_{t,b,\bar{\beta}} \left\{ \eta^+(\mathbf{y})_{tb\bar{\beta}}\, \eta(\mathbf{y})_{tb\bar{\beta}} + \zeta^+(\mathbf{y})_{tb\bar{\beta}}\, \zeta(\mathbf{y})_{tb\bar{\beta}} \right\}\right] = 1\,. \tag{4.18}$$

We are now in the position to express any Fock space operator $\hat{T}$ as integral operator in the Grassmann algebra by applying the isomorphism to the state $\hat{T}|X\rangle$ and inserting the identity once

$$\langle\eta^+\zeta^+|\hat{T}|X\rangle = \int \mathcal{D}[\chi']\mathcal{D}[\bar{\chi}']\,\mathrm{e}^{-\bar{\chi}'\chi'}\langle\eta^+\zeta^+|\hat{T}|\eta'\,\zeta'\rangle\langle\eta'^+\zeta'^+|X\rangle. \tag{4.19}$$

The measure appearing here is the same as in eq.(4.17) (cf. the remark after eq.(4.13)) and defined through

$$\mathcal{D}[\chi]\mathcal{D}[\bar{\chi}] = \prod_{\mathbf{x}\in\Gamma}\prod_{s,a,\alpha} \mathrm{d}\chi(\mathbf{x})_{sa\alpha}\mathrm{d}\bar{\chi}(\mathbf{x})_{sa\alpha}. \tag{4.20}$$

Products and powers of operators can be treated similarly by inserting the identity operation (4.17) between any two factors.

### 4.3 Grassmann representation of the transfer matrix

All one has to do now, is to apply this procedure to the transfer matrix (4.6) and its $N$-th power. Starting from equation (4.7), one first expresses the operator $\hat{T}_F^\dagger(U)\hat{T}_F(U')$ in the two-component spinors and reorders it such as to have the crossed operators to the left of the uncrossed operators,

$$\begin{aligned}\hat{T}_F^\dagger(U)\hat{T}_F(U') = {}& \det(2\mathcal{K}B)^{1/4}\det(2\mathcal{K}B')^{1/4} \\ & \times \exp(\hat{\eta}^\dagger e^{-\bar{M}}\bar{C}e^{-\bar{M}}\hat{\zeta}^\dagger)\exp(-\hat{\eta}^\dagger \bar{M}\hat{\eta} + \hat{\zeta}\bar{M}\hat{\zeta}^\dagger) \\ & \times \exp(-\hat{\eta}^\dagger \bar{M}'\hat{\eta} + \hat{\zeta}\bar{M}'\hat{\zeta}^\dagger)\exp(\hat{\zeta}e^{-\bar{M}'}\bar{C}'e^{-\bar{M}'}\hat{\eta}).\end{aligned} \qquad (4.21)$$

The corresponding integral kernel with respect to the Grassmann variables is then found to be

$$\begin{aligned}\langle\eta^+\zeta^+|\hat{T}_F^\dagger(U)\hat{T}_F(U')|\eta\zeta\rangle = {}& \det(BB')^{1/2}\exp(\eta^+\bar{B}^{-1/2}2\mathcal{K}\bar{C}\bar{B}^{-1/2}\zeta^+) \\ & \times \exp(\eta^+\bar{B}^{-1/2}2\mathcal{K}\bar{B}'^{-1/2}\eta - \zeta\bar{B}'^{-1/2}2\mathcal{K}\bar{B}^{-1/2}\zeta^+) \qquad (4.22)\\ & \times \exp(\zeta\bar{B}'^{-1/2}2\mathcal{K}\bar{C}'\bar{B}'^{-1/2}\eta).\end{aligned}$$

By definition the crossed Grassmann variables $\eta^+$ and $\zeta^+$ correspond to $\bar{\chi}P_+ \equiv \bar{\chi}_+$ and $P_-\chi \equiv \chi_-$ [cf. eq.(4.13)]. Since the matrix $B$ commutes with the projectors $P_\pm$, one may also define a wave function $X$ as a function of the components $\psi_-$ and $\bar{\psi}_+$, viz.

$$X(\bar{\psi}_+,\psi_-,U) \stackrel{\text{def}}{=} \langle\eta^+\zeta^+|X(U)\rangle. \qquad (4.23)$$

Performing the change of variables in the Grassmann integral, using eq.(4.22) and the transformation of the measure (4.20), $\mathcal{D}[\chi]\mathcal{D}[\bar{\chi}] = \mathcal{D}[\psi]\mathcal{D}[\bar{\psi}]/\det B$, one finds the full action of the operator $\mathbb{T}_0$ on a wave function

$$\begin{aligned}(\mathbb{T}_0 X)(\bar{\psi}_+,\psi_-,U) = {}& \int \mathcal{D}[U']\mathcal{D}[\psi']\mathcal{D}[\bar{\psi}']/\det B'\, e^{-\bar{\psi}'B'\psi'}\det(BB')^{1/2} \\ & \times e^{-2\{\bar{\psi}P_+\mathcal{K}C\psi - \bar{\psi}P_+\mathcal{K}\psi' - \bar{\psi}'P_-\mathcal{K}\psi + \bar{\psi}'P_-\mathcal{K}C'\psi'\}} \qquad (4.24)\\ & \times e^{-\Delta S[U,1,U']} X(\bar{\psi}'_+,\psi'_-,U').\end{aligned}$$

As for the pure gauge theory, one introduces the projector $\mathbb{P}$ on the physical subspace of gauge invariant wave functions,

$$\mathbb{P}X(\bar{\psi}_+,\psi_-,U) = \int \prod_{\mathbf{x}\in\Gamma}\mathrm{d}V(\mathbf{x})\, X(\bar{\psi}_+ V^{-1},V\psi_-,U^V). \qquad (4.25)$$

The operator $\mathbb{T}_0$ is again referred to as transfer matrix in the temporal gauge. The full integral kernel of the transfer matrix $\mathbb{T} = \mathbb{T}_0 \mathbb{P} = \mathbb{P} \mathbb{T}_0$ in terms of the classical four-component spinors is now easily found, making use of the gauge invariance of the measure

$$\langle \eta^+ \zeta^+ | \mathbb{T}[U, U'] | \eta' \zeta' \rangle = \det(BB')^{1/2} \int \prod_{\mathbf{x} \in \Gamma} dV(\mathbf{x})\, e^{-\Delta S[U, V, U']} \tag{4.26}$$
$$\times e^{-2\{\bar{\psi} P_+ \mathcal{K} C \psi - \bar{\psi} P_+ \mathcal{K} V^{-1} \psi' - \bar{\psi}' V P_- \mathcal{K} \psi + \bar{\psi}' P_- \mathcal{K} C \psi'\}}.$$

*4.4 Definition of the Schrödinger functional*

It is not difficult to iterate the procedure and to evaluate the $N$-th power of the transfer matrix. Its kernel

$$\langle \eta_N^+ \zeta_N^+ | \mathbb{T}^N[U_N, U_0] | \eta_0 \zeta_0 \rangle = \int \prod_{l=1}^{N-1} \mathcal{D}[U_l] \mathcal{D}[\chi_l] \mathcal{D}[\bar{\chi}_l]\, e^{-\bar{\chi}_l \chi_l} \tag{4.27}$$
$$\times \langle \eta_N^+ \zeta_N^+ | \mathbb{T}[U_N, U_{N-1}] | \eta_{N-1} \zeta_{N-1} \rangle \cdots$$
$$\cdots \langle \eta_2^+ \zeta_2^+ | \mathbb{T}[U_2, U_1] | \eta_1 \zeta_1 \rangle \langle \eta_1^+ \zeta_1^+ | \mathbb{T}[U_1, U_0] | \eta_0 \zeta_0 \rangle$$

can be obtained in terms of the classical quark fields $\psi$ and $\bar{\psi}$ by inserting the expression (4.26) in the r.h.s. of this equation and changing the measure accordingly. One then obtains the following expression for the kernel of $\mathbb{T}^N$

$$\det(B_N B_0)^{1/2} \int \prod_{\mathbf{x}} dV_0(\mathbf{x}) \int \prod_{l=1}^{N-1} \mathcal{D}[U_l]\mathcal{D}[\psi_l]\mathcal{D}[\bar{\psi}_l] \prod_{\mathbf{x}} dV_l(\mathbf{x})\, e^{-\bar{\psi}_l B_l \psi_l}$$
$$\times \exp - \sum_{k=0}^{N-1} \Big\{ \Delta S[U_{k+1}, V_k, U_k] + \bar{\psi}_k P_- 2\mathcal{K} C_k \psi_k - \bar{\psi}_k P_- 2\mathcal{K} V_k \psi_{k+1} \tag{4.28}$$
$$- \bar{\psi}_{k+1} P_+ 2\mathcal{K} V_k^{-1} \psi_k + \bar{\psi}_{k+1} P_+ 2\mathcal{K} C_{k+1} \psi_{k+1} \Big\}.$$

As announced above, the label $l$ is interpreted as euclidean time, i.e. one sets

$$l = x^0/a, \quad N = T/a. \tag{4.29}$$

The link variables $U_l(\mathbf{x}, k)$ and the fermi fields $\psi_l(\mathbf{x})$, $\bar{\psi}_l(\mathbf{x})$ are then identified accordingly with $U(x, k)$, $\psi(x)$ and $\bar{\psi}(x)$. The fields $V_l$, stemming from the projectors $\mathbb{P}$ play the rôle of the time-like link variables $V_l(\mathbf{x}) = U(x, 0)$, for $l = 0, .., N-1$. Up to a normalization factor, the kernel (4.28) is the Schrödinger functional we have been looking for. More precisely, we define

$$\mathcal{Z} = \det(B_N B_0)^{-1/2} \langle \eta_N^+ \zeta_N^+ | \mathbb{T}^N[U_N, U_0] | \eta_0 \zeta_0 \rangle, \tag{4.30}$$

thus discarding the determinants of the gauge field at the boundary. We are free to do so, because the boundary configurations are not integrated over. After all, the determinants have been introduced in order to cancel the Jacobians when the formula for the partition function is derived [4].

One may consider the Schrödinger functional as a functional of the fields at the boundary. In view of the continuum limit, we will now use the rescaled fields $\psi_c = (2\mathcal{K})^{1/2}\psi$ (cf. sect.1) with boundary values denoted as follows

$$
\begin{aligned}
P_+\psi_c(x)|_{x_0=0} = \rho_+(\mathbf{x}), \qquad & P_-\psi_c(x)|_{x_0=T} = \rho'_-(\mathbf{x}), \\
\bar{\psi}_c(x)P_-|_{x_0=0} = \bar{\rho}_-(\mathbf{x}), \qquad & \bar{\psi}_c(x)P_+|_{x_0=T} = \bar{\rho}'_+(\mathbf{x}).
\end{aligned}
\tag{4.31}
$$

The generalization of the pure gauge Schrödinger functional (3.15) has the path integral representation

$$
\mathcal{Z}[\bar{\rho}'_+, \rho'_-, C'; \bar{\rho}_-, \rho_+, C] = \int \mathrm{D}[\psi]\mathrm{D}[\bar{\psi}]\mathrm{D}[U]\, \mathrm{e}^{-S}, \tag{4.32}
$$

where the integration is over all quark fields at euclidean times $x_0$ with $0 < x_0 < T$ and over the euclidean gauge field as specified at the end of section 3. The gauge part $S_g$ of the action $S = S_g + S_f$ has also been discussed there and the fermionic part $S_f$ is given by

$$
\begin{aligned}
S_f = \; & a^3 \sum_{\mathbf{x}\in\Gamma} \bar{\psi}_c(x)P_-\Big\{\sum_{k=1}^{3} a\gamma_k\widetilde{\nabla}_k\psi_c(x) - U(x,0)\psi_c(x+a\hat{0})\Big\}_{x_0=0} \\
& + a^4 \sum_{\mathbf{x}\in\Gamma} \sum_{x^0=a}^{T-a} \bar{\psi}_c(x)\Big\{\sum_\mu \widetilde{\nabla}_\mu\gamma_\mu + m + \frac{1}{2}a\Delta\Big\}\psi_c(x) \\
& + a^3 \sum_{\mathbf{x}\in\Gamma} \bar{\psi}_c(x)P_+\Big\{\sum_{k=1}^{3} a\gamma_k\widetilde{\nabla}_k\psi_c(x) - U(x-a\hat{0},0)^{-1}\psi_c(x-a\hat{0})\Big\}_{x_0=T}.
\end{aligned}
\tag{4.33}
$$

Apart from the boundary terms, this is the Wilson action (2.6), again expressed in continuum notation. It should be emphasized that the whole boundary dependence is through the field components which fulfill the inhomogeneous Dirichlet conditions (4.31), the complementary components being absent. This can be understood from our derivation, since any Grassmann integral kernel only depends of $\eta$ and $\zeta$ at the former and $\eta^+$ and $\zeta^+$ at the later euclidean time. It can, of course, also be verified directly by inspection of the action (4.33). While this is easy to see as long as the lattice is present, it would be more difficult to establish in the continuum, because the difference between the fields at and near the boundary vanishes.

# 5. Continuum limit of the action

As a first step in the investigation of the Schrödinger functional (4.32), one would like to know, whether it leads to a sensible classical theory in the continuum. In this section we thus consider the corresponding lattice action in the limit of vanishing lattice spacing $a$. To this end a gauge potential $A_\mu$ is introduced by parameterizing the link variables as follows

$$U(x,\mu) = e^{aA_\mu(x)}.\tag{5.1}$$

The classical continuum action $S^c$ is then obtained as the first term in the small $a$-expansion

$$S[U,\bar\psi,\psi] = S^c[A,\bar\psi,\psi] + \mathcal{O}(a).\tag{5.2}$$

For the fermionic part of the continuum action, one finds

$$S_f^c[A,\bar\psi,\psi] = \int_0^T dx_0 \int_0^L d^3\mathbf{x}\, \bar\psi(x)\left\{\gamma_\mu D_\mu + m\right\}\psi(x) \\ - \int_0^L d^3\mathbf{x}\,\left[\bar\psi(x)P_-\psi(x)\right]_{x_0=0} - \int_0^L d^3\mathbf{x}\,\left[\bar\psi(x)P_+\psi(x)\right]_{x_0=T}.\tag{5.3}$$

Here, the covariant derivative is defined by $D_\mu = \partial_\mu + A_\mu$, and the projectors $P_\pm = \frac{1}{2}(1\pm\gamma_0)$ are the same as in the preceding sections. The quark fields are $L$-periodic in space and subject to the boundary conditions

$$\begin{aligned} P_+\psi(x)|_{x_0=0} &= \rho_+(\mathbf{x}), & P_-\psi(x)|_{x_0=T} &= \rho'_-(\mathbf{x}), \\ \bar\psi(x)P_-|_{x_0=0} &= \bar\rho_-(\mathbf{x}), & \bar\psi(x)P_+|_{x_0=T} &= \bar\rho'_+(\mathbf{x}). \end{aligned}\tag{5.4}$$

At first sight, the presence of the boundary terms in the action (5.3) is surprising. One might have expected to obtain the usual euclidean action, with the boundary conditions (5.4) imposed on the fields. At least, this is what happens in the pure gauge theory and the reader might be worried whether the boundary terms are a remnant of the lattice formulation. Stated differently, the question arises, whether it is possible to understand the boundary terms without reference to the lattice.

To answer the question, one must first make precise what exactly one means by a sensible classical continuum theory. It will be assumed here, that the classical action is a functional on the space of smooth $C^\infty$-functions, which satisfy the boundary conditions (5.4). One may then set up a variational principle and look for the stationary points of the action. We will require the existence of smooth stationary points, which means that the corresponding solutions are required to be again in the above space of functions.

Using these assumptions, one may investigate a fermionic continuum action with a general ansatz for the boundary terms,

$$S_f[A, \bar\psi, \psi] = \int_0^T dx_0 \int_0^L d^3\mathbf{x}\, \bar\psi(x)\{\gamma_\mu D_\mu + m\}\psi(x)$$
$$+ \int_0^L d^3\mathbf{x}\, [\bar\psi(x) R_0(\mathbf{x})\psi(x)]_{x_0=0} + \int_0^L d^3\mathbf{x}\, [\bar\psi(x) R_T(\mathbf{x})\psi(x)]_{x_0=T}\,. \quad (5.5)$$

The matrices $R_0(\mathbf{x})$ and $R_T(\mathbf{x})$ parameterize the boundary terms and are left arbitrary for the moment.

The variations $\delta S$ and $\bar\delta S$, defined through

$$\delta S(v) = \lim_{\epsilon\searrow 0} \frac{S[A,\psi+\epsilon v,\bar\psi] - S[A,\psi,\bar\psi]}{\epsilon},$$
$$\bar\delta S(\bar v) = \lim_{\epsilon\searrow 0} \frac{S[A,\psi,\bar\psi+\epsilon\bar v] - S[A,\psi,\bar\psi]}{\epsilon}, \quad (5.6)$$

are distributions, evaluated on smooth test functions $v$ and $\bar v$. The test functions have to respect the boundary conditions on $\psi$ and $\bar\psi$, i.e. they are L-periodic in the spatial directions and satisfy homogenous boundary conditions

$$P_+ v(x)|_{x_0=0} = 0, \qquad P_- v(x)|_{x_0=T} = 0,$$
$$\bar v(x) P_-|_{x_0=0} = 0, \qquad \bar v(x) P_+|_{x_0=T} = 0\,. \quad (5.7)$$

One then finds

$$\bar\delta S(\bar v) = \int_0^T dx_0 \int_0^L d^3\mathbf{x}\, \bar v(x)\{\gamma_\mu D_\mu + m + \delta(x_0) P_+ R_0(\mathbf{x}) + \delta(x_0-T) P_- R_T(\mathbf{x})\}\psi(x), \quad (5.8)$$

and, after integration by parts

$$\delta S(v) = \int_0^T dx_0 \int_0^L d^3\mathbf{x}\, \bar\psi(x)\{\gamma_\mu[-\overleftarrow{\partial}_\mu + A_\mu(x)] + m$$
$$+ \delta(x_0)(R_0(\mathbf{x}) - \gamma_0)P_- + \delta(x_0-T)(R_T(\mathbf{x}) + \gamma_0)P_+\}v(x)\,. \quad (5.9)$$

According to the assumptions stated above, one requires the existence of smooth solutions $\psi_{cl}$ and $\bar\psi_{cl}$ of the equations

$$\delta S(v) = 0, \qquad \bar\delta S(\bar v) = 0. \quad (5.10)$$

From eq.(5.8), one infers that $\psi_{cl}$ must be a solution of $(\gamma_\mu D_\mu + m)\psi_{cl}(x) = 0$ for all points $x$ with $0 < x_0 < T$. The assumption of smoothness then implies that this must also be the case at the points $x_0 = 0$ and $x_0 = T$, i.e. one must have

$$P_+ R_0(\mathbf{x})\psi_{cl}(x)|_{x_0=0} = 0, \qquad P_- R_T(\mathbf{x})\psi_{cl}(x)|_{x_0=T} = 0\,. \quad (5.11)$$

The same arguments, applied to $\bar{\psi}_{cl}$ lead to

$$\bar{\psi}_{cl}(x)(R_0(\mathbf{x}) - \gamma_0)P_-|_{x_0=0} = 0, \qquad \bar{\psi}_{cl}(x)(R_T(\mathbf{x}) + \gamma_0)P_+|_{x_0=T} = 0. \tag{5.12}$$

Since one does not want to impose further conditions on the fields at the boundary, these equations are really conditions on the matrices $R_0(\mathbf{x})$ and $R_T(\mathbf{x})$, which must satisfy

$$\begin{aligned} P_+ R_0(\mathbf{x}) &= 0, & (R_0(\mathbf{x}) - \gamma_0)P_- &= 0, \\ P_- R_T(\mathbf{x}) &= 0, & (R_T(\mathbf{x}) + \gamma_0)P_+ &= 0. \end{aligned} \tag{5.13}$$

The solutions are

$$R_0(\mathbf{x}) = -P_- + P_-\Gamma_0(\mathbf{x})P_+, \qquad R_T(\mathbf{x}) = -P_+ + P_+\Gamma_T(\mathbf{x})P_-. \tag{5.14}$$

Here, $\Gamma_0(\mathbf{x})$ and $\Gamma_T(\mathbf{x})$ are again arbitrary matrices. However, to give a non-vanishing contribution to the action (5.5), they have to anti-commute with $\gamma_0$. Such matrices are ruled out if one requires the action to be invariant under parity.

In conclusion, one may say that the general ansatz (5.5) coincides with the fermionic action (5.3) as derived from the Schrödinger functional, provided the existence of smooth stationary points and parity invariance of the action are required. This is nice, because no reference to the lattice is needed, once the boundary conditions (5.4) are known.

## 6. The free Dirac operator

In the preceding section it has been shown how the action (5.3) can be understood without reference to the lattice formulation. An essential ingredient was the existence of a smooth stationary point, which implies that the decomposition

$$S^c[A, \bar{\psi}_{cl} + \bar{v}, \psi_{cl} + v] = S^c[A, \bar{\psi}_{cl}, \psi_{cl}] + S^c[A, \bar{v}, v], \tag{6.1}$$

does not contain linear terms in the fluctuation fields $\bar{v}$ and $v$. To get acquainted with the boundary conditions (5.7), it is helpful to have a closer look at the mathematical structure involved. Starting from the free action $S_f^c[\bar{v}, v]$, we first define the free Dirac operator. Motivated by potential applications we then establish its essential self-adjointness and determine its spectrum.

## 6.1 Definition of the free Dirac operator

As usual, one would like to write the action $S_f^c[\bar{v}, v]$ as a quadratic form in a suitable linear space of functions. At this point we observe that the boundary conditions (5.7) on $v$ and $\bar{v}$ allow for the definition of two different linear spaces of smooth, spatially periodic functions $u(x)$, namely

$$\mathcal{H}_\pm \stackrel{\text{def}}{=} \left\{ u \,\middle|\, P_\pm u(x)\big|_{x_0=0} = 0, \ P_\mp u(x)\big|_{x_0=T} = 0 \right\}. \tag{6.2}$$

Equipped with the inner product

$$(u_1, u_2)_{\mathcal{H}_\pm} \stackrel{\text{def}}{=} \int_0^T dx_0 \int_0^L d^3\mathbf{x} \sum_{\alpha=1}^4 u_1^*(x)_\alpha \, u_2(x)_\alpha \,, \tag{6.3}$$

$\mathcal{H}_\pm$ are pre-Hilbert spaces. A smooth spinor field $v$ is an element of $\mathcal{H}_+$, while $\bar{v}$ defines an element $\bar{v}^\dagger \in \mathcal{H}_-$. The free fermionic action can be written as a quadratic form in either $\mathcal{H}_+$ or $\mathcal{H}_-$,

$$S_f^c[\bar{v}, v] = (\bar{v}^\dagger, Dv)_{\mathcal{H}_-} = (D^\dagger \bar{v}^\dagger, v)_{\mathcal{H}_+}, \tag{6.4}$$

where it is assumed that the operator $D = \gamma_\mu \partial_\mu + m$ and its formal adjoint $D^\dagger$ act as follows

$$D : \mathcal{H}_+ \mapsto \mathcal{H}_-, \qquad D^\dagger : \mathcal{H}_- \mapsto \mathcal{H}_+ \,. \tag{6.5}$$

Since both operators are elliptic (their leading symbols $\pm i p_\mu \gamma_\mu$ are invertible for $p_\mu \neq 0$), this defines a structure, which is known in the mathematical literature as a two-term elliptic complex (cf. ref.[8]).

We note that the eigenvalue problem for the operator $D$ is not well-defined. However, according to Gilkey [9], one can obtain a first order elliptic boundary value problem if one defines an operator $\mathcal{D}$ as follows,

$$\mathcal{D} : \mathcal{H}_+ \oplus \mathcal{H}_- \mapsto \mathcal{H}_+ \oplus \mathcal{H}_-, \qquad \mathcal{D} \stackrel{\text{def}}{=} \begin{pmatrix} 0 & D^\dagger \\ D & 0 \end{pmatrix}. \tag{6.6}$$

The operator $\mathcal{D}$ thus acts on two-component vectors with the upper and lower components being elements of $\mathcal{H}_+$ and $\mathcal{H}_-$ respectively. It is an elliptic operator which is also hermitian with respect to the induced inner product. We will refer to it as the Dirac operator in the following.

A further notational item are the projectors

$$\mathcal{P}_\pm = \begin{pmatrix} P_\pm & 0 \\ 0 & P_\mp \end{pmatrix}, \tag{6.7}$$

which allow to characterize the linear space $\mathcal{H}_+ \oplus \mathcal{H}_-$ as follows

$$\mathcal{H} \stackrel{\text{def}}{=} \mathcal{H}_+ \oplus \mathcal{H}_- = \left\{ w \,\middle|\, \mathcal{P}_+ w(x)\big|_{x_0=0} = 0, \ \mathcal{P}_- w(x)\big|_{x_0=T} = 0 \right\}. \tag{6.8}$$

The requirement that $\mathcal{D}w$ be an element of $\mathcal{H}$ entails the boundary conditions for the complementary components

$$(\partial_0 - m)\mathcal{P}_-w(x)|_{x_0=0} = 0, \qquad (\partial_0 + m)\mathcal{P}_+w(x)|_{x_0=T} = 0. \tag{6.9}$$

With these additional boundary conditions, the Dirac operator squared,

$$\mathcal{D}^2 = \begin{pmatrix} D^\dagger D & 0 \\ 0 & DD^\dagger \end{pmatrix}, \tag{6.10}$$

also defines an elliptic boundary value problem. As will be shown below, the Dirac operator $\mathcal{D}$ and its square have unique self-adjoint extensions with purely discrete spectrum and a complete set of smooth eigenfunctions.

*6.2 Relevant questions*

If one wants to carry out a perturbative expansion of the Schrödinger functional, it is important to know whether the free quark propagator, i.e. the operator $D^{-1}$ is well-defined. As will be shown below, the operator $\mathcal{D}^2$ is essentially self-adjoint and strictly positive for any value of the mass. Eq.(6.10) then implies that this is also true for $DD^\dagger$ and $D^\dagger D$. It is therefore possible to define the propagator as follows

$$D^{-1} = D^\dagger (DD^\dagger)^{-1}. \tag{6.11}$$

In case dimensional regularization is employed, it is important to note that this statement remains valid for the $d$-dimensional theory, provided translation invariance is not violated in the additional dimensions.

Ultimately we will be interested in numerical simulations on the lattice. Current fermion algorithms involve the inversion of the matrix $D_{\text{Lat}}^\dagger D_{\text{Lat}}$, i.e. the lattice version of $D^\dagger D$. The convergence rate of the matrix inversion is then determined by the so-called condition number, which is the ratio of the largest to the smallest eigenvalue [10]. Therefore, one has to make sure that the smallest eigenvalue is not too close to zero. For simulations in large volumes, this means that it is difficult to reach very light quark masses. The situation is different in small volumes, where the influence of the boundary conditions becomes important. In this case, one could envisage the simulation of light or even massless quarks, provided a smallest eigenvalue is induced by the boundary conditions. As will be explained in section 7, such simulations are required for the determination of the strong coupling at high energies, using the strategy of refs.[1,2].

We are thus led to investigate the continuum operator $\mathcal{D}^2$ (6.10) and its spectrum in some detail. Close to the continuum limit, one expects that its lowest eigenvalues are well approximated on the lattice. It may be anticipated here that the boundary conditions (5.7) show their quality in inducing a minimal eigenvalue, given by $(\pi/2T)^2$ in the massless case.

## 6.3 Self-adjointness and spectrum of $\mathcal{D}^2$

The operator $\mathcal{D}^2$ is defined on functions $w \in \mathcal{H}$ which satisfy the additional boundary condition (6.9). A considerable simplification is achieved by noting that, owing to spatial periodicity, $\mathcal{D}^2$ is invariant under spatial translations. The eigenfunctions of this transformation group are the plane waves $e^{i\mathbf{p}\mathbf{x}}$, where the components $p_k$ of the spatial momentum $\mathbf{p}$ assume the values $2\pi n_k/L$, with integer numbers $n_k$. The corresponding change of basis is given by the Fourier transformation

$$w(x) = \frac{1}{L^3} \sum_{\mathbf{p}} e^{i\mathbf{p}\mathbf{x}} w_{\mathbf{p}}(x_0), \qquad w_{\mathbf{p}}(x_0) = \int_0^L d^3\mathbf{x}\, e^{-i\mathbf{p}\mathbf{x}} w(x). \qquad (6.12)$$

In the following we restrict attention to the subspaces of definite momentum $\mathbf{p}$, where the spatial derivative $\partial_k$ amounts to a multiplication by $ip_k$.

The operator $\mathcal{D}^2$ is diagonal in Dirac space, so we may consider its action on each spinor component separately. A spinor component $w_{\mathbf{p}}(x_0)_l$ is an ordinary function of $x_0$ which will be denoted by $s(x_0)$ in the following. One is thus led to study the differential operator $-\partial_0^2 + \mathbf{p}^2 + m^2$ on smooth functions $s(x_0)$ which satisfy either

$$s(0) = 0, \quad s'(T) + ms(T) = 0, \quad \text{or} \quad s(T) = 0, \quad s'(0) - ms(0) = 0, \qquad (6.13)$$

depending on whether $s(x_0)$ belongs to the subspace defined by the projector $\mathcal{P}_+$ (plus-components) or by $\mathcal{P}_-$ (minus-components). From the literature (see e.g. [11], p. 237), one infers that this operator is essentially self-adjoint in either case. Moreover, its spectrum is purely discrete, all eigenvalues are non-degenerate and one may find an orthonormal basis consisting of its eigenfunctions.

To find the eigenvalues and eigenfunctions, we write the corresponding eigenvalue equation

$$(-\partial_0^2 + \mathbf{p}^2 + m^2)s = \lambda^2 s \qquad (6.14)$$

in the form

$$s'' + p_0^2 s = 0, \qquad p_0^2 = \lambda^2 - \mathbf{p}^2 - m^2, \qquad (6.15)$$

which is simply the differential equation of a harmonic oscillator with frequency $p_0$. Taking into account the boundary conditions, the eigenfunctions are found to be of the form $\sin p_0 x_0$ for the plus-components and $\sin p_0(x_0 - T)$ for the minus-components. In either case, one determines the allowed values for $p_0$ as the solutions of the equation

$$\tan p_0 T = -\frac{p_0}{m}. \qquad (6.16)$$

The case $p_0 = 0$ is special. The boundary conditions exclude a non-vanishing solution of $s'' = 0$, except for the particular value of the mass $m = -1/T$. In this case, there are linear eigenfunctions, which are proportional to $x_0$ (plus-components) or $x_0 - T$ (minus-components).

Depending on the value of $m$, one has to distinguish 3 different cases for the solutions of eq.(6.16). For non-negative masses, all solutions are real. Since the eigenfunctions are odd functions of $p_0$ one may restrict attention to the positive solutions. They lie in the intervals

$$p_0(n_0) \in \left[ \frac{(n_0 + \frac{1}{2})\pi}{T}, \frac{(n_0 + 1)\pi}{T} \right), \qquad n_0 = 0, 1, 2, \ldots \tag{6.17}$$

For vanishing mass, the solutions are then given as the left limits of the intervals in (6.17), with $\pi/2T$ as the lowest possible value.

The second case corresponds to masses in the range $0 > m \geq -1/T$. The solutions are again real and labelled as follows

$$p_0(n_0) \in \left[ \frac{n_0 \pi}{T}, \frac{(n_0 + \frac{1}{2})\pi}{T} \right), \qquad n_0 = 0, 1, 2, \ldots \tag{6.18}$$

In the last case, the mass is negative and below $-1/T$. Equation (6.16) admits real solutions in the same intervals as in (6.18) for $n_0 > 0$. In addition, one finds a purely imaginary solution $p_0(0) = iq, q > 0$. One may continue to write e.g. $\sin p_0 x_0$ for the corresponding eigenfunction, but one should be aware that this corresponds to the hyperbolic sine function $i \sinh q x_0$.

With these conventions, the complete set of orthonormal eigenfunctions of $\mathcal{D}^2$ is given by $\{\varphi_{n,l}(x)\}$, with

$$\varphi_{n,l}(x) = \begin{cases} \sqrt{3} \, (LT)^{-3/2} \, e^{i\mathbf{p}\mathbf{x}} [\mathcal{P}_+ x_0 + \mathcal{P}_-(x_0 - T)] e_l, & \text{if } n_0 = 0, \, m = -1/T, \\ N(p_0) \, e^{i\mathbf{p}\mathbf{x}} [\mathcal{P}_+ \sin p_0 x_0 + \mathcal{P}_- \sin p_0(x_0 - T)] e_l, & \text{otherwise,} \end{cases}$$

$$N(p_0) = \sqrt{\frac{2}{L^3 T}} \left( 1 + \frac{m/T}{p_0^2 + m^2} \right)^{-1/2}.$$
(6.19)

Here, $e_l$, $l = 1, .., 8$ is a canonical basis in $\mathbb{C}^8$. The components $n_\mu$ run over all integer numbers if $\mu = 1, 2, 3$, labelling the spatial momenta $\mathbf{p} = 2\pi \mathbf{n}/L$, and over all non-negative integer numbers if $\mu = 0$, labelling the zero component $p_0$ according to eqs.(6.17) or (6.18).

The corresponding eigenvalues $\lambda_n^2$ are given by

$$\lambda_n^2 = p_0^2 + \mathbf{p}^2 + m^2, \tag{6.20}$$

each of them defining an 8-dimensional eigenspace. As suggested by the notation, the eigenvalues are all non-negative. The condition for the appearance of a zero mode may be derived from eq.(6.16), by substituting for $p_0$ (6.15) and setting $\lambda$ to zero

$$\tanh \sqrt{\mathbf{p}^2 + m^2} T = -\frac{\sqrt{\mathbf{p}^2 + m^2}}{m}. \tag{6.21}$$

The l.h.s. of this equation assumes values between 0 and 1, while the r.h.s. is below $-1$ for positive masses and above 1 for negative values of $m$. A zero mode is thus obtained only in the limit of infinite negative mass.

We may summarize the result obtained so far by saying that the square of the Dirac operator is an essentially self-adjoint, positive operator for all values of the mass. Its spectrum is purely discrete, with the lowest eigenvalue given by $(\pi/2T)^2$ in the massless case. Furthermore, its eigenfunctions (6.19) are smooth and define an orthonormal basis in the Hilbert space (i.e. the completion of $\mathcal{H}$).

Concerning the Dirac operator itself, we note that its action on the eigenfunctions of $\mathcal{D}^2$ reduces to the action of a hermitian matrix in each subspace of fixed $n$. One has

$$\mathcal{D} \sum_{l=1}^{8} w_l \varphi_{n,l}(x) = \sum_{l=1}^{8} w'_l \varphi_{n,l}(x), \tag{6.22}$$

where $w' \in \mathbb{C}^8$ is given by $w' = \mathcal{A}w$, with the matrix $\mathcal{A}$ defined as follows

$$\mathcal{A} = \begin{pmatrix} 0 & A^\dagger \\ A & 0 \end{pmatrix}, \qquad A \stackrel{\text{def}}{=} ip_k \gamma_k + (-1)^{n_0+1} \sqrt{p_0^2 + m^2}. \tag{6.23}$$

This is a hermitian matrix with real eigenvalues, which are determined as the roots of $\lambda_n^2$ (6.20),

$$\lambda_{n,\pm} = \pm\sqrt{p_0^2 + \mathbf{p}^2 + m^2}. \tag{6.24}$$

Furthermore, since the trace of $\mathcal{A}$ vanishes, each eigenvalue is fourfold degenerate.

Except for positivity, the operator $\mathcal{D}$ has the same properties as its square. In particular, the set of functions (6.19) defines a core [12] which determines a unique self-adjoint extension of the Dirac operator.

## 7. Concluding remarks

So far, we have given a definition of the Schrödinger functional in lattice QCD. The formal continuum limit of the corresponding lattice action led to an acceptable classical continuum action, with a well-defined free Dirac operator. In this last section, we briefly discuss the renormalizability of the Schrödinger functional and comment on its application to the determination of the strong coupling constant.

*7.1 Renormalizability of the Schrödinger functional*

The Schrödinger functional naturally leads to a field theory with prescribed boundary values for some of the field components. In general, quantum field theories with boundaries are afflicted by additional divergences. It is then a priori not clear whether the theory stays renormalizable.

In his work on the Schrödinger representation in quantum field theory, Symanzik considered the case of massless scalar $\phi^4$ field theory [13] (for an introduction to this paper see also ref.[14]). He found that the Schrödinger functional could be made finite by adding two new counterterms to the action. These are proportional to the local composite fields $\phi\partial_0\phi$ and $\phi^2$, integrated over the hyper-planes at $x_0 = 0$ and $x_0 = T$.

Following Symanzik, it is plausible that this behaviour is generic for a renormalizable quantum field theory. We thus have to look for possible boundary counterterms in the case of QCD. If we require invariance under parity, there are only two gauge invariant local composite fields of dimension 3 or less, namely

$$\bar\psi\psi, \qquad \text{and} \qquad \bar\psi\gamma_0\psi. \tag{7.1}$$

We therefore expect that the QCD Schrödinger functional is renormalizable after inclusion of at most two boundary counterterms, which are linear combinations of the composite fields in eq.(7.1). To confirm this expectation, it would be desirable to carry through a one-loop calculation with the QCD Schrödinger functional. This is currently undertaken in the framework of dimensional regularization, with the action (5.3) as starting point [15].

*7.2 Application to the determination of $\alpha_s$*

It is our aim to apply the formalism to the determination of the strong coupling at high energies, following the strategy of refs.[1,2]. For the gauge fields, one could take over the formalism as developed for the the pure $SU(3)$ gauge theory. In particular one could choose the same boundary conditions and the same definition of the running coupling. Concerning the quark fields, one would be tempted to set the boundary fields (4.31) to zero. If the renormalizability of the Schrödinger functional works out as expected, this implies the vanishing of the boundary counterterms [cf. eq.(7.1)]. The effect of the boundary is then of the order $a$ and will thus modify the cutoff dependence of the theory.

The Schrödinger functional serves to define a coupling which is assumed to run with $L$, the linear extent of the volume (one sets $T = L$). In order to trace its evolution from low to high energies, one has to cover a corresponding range of volumes.

As has been shown in the preceding section, the boundary conditions force a minimal frequency into the system. In small volumes, or, equivalently, at high energies one therefore has a chance to simulate QCD with light quarks. As a first step, one could then try to simulate QCD with only massless quarks, thus staying with the bare coupling as a single free parameter. Of course, this is not quite true on the lattice with Wilson fermions, where chiral symmetry is broken explicitly. In principle, one must carefully tune the bare quark masses in order to find the chiral limit. However, it is plausible that the coupling at high energies is not very sensitive to the precise values of the light quark masses, so that effectively only 1 parameter, the bare gauge coupling has to be tuned.

It is difficult to overestimate the importance of a minimal frequency induced by the boundary. Otherwise, one would have to keep the quark masses at relatively large values so that a clean determination of the strong coupling might not be feasible.

In sect.6, only the free theory has been discussed. Of course, the gauge interactions will modify the spectrum of the Dirac operator and one expects the appearance of zero modes, corresponding to some particular gauge configurations. It will then be necessary to choose the background gauge field such that these gauge configurations become extremely unlikely. This will give an additional criterion for a good choice of boundary conditions for the gauge fields. At present it is not clear whether this criterion is satisfied by the background field which has been used for the simulations of the $SU(3)$ pure gauge theory.

I would like to thank M. Lüscher for advice and many helpful discussions.